\documentstyle[prd,preprint,tighten,aps,eqsecnum,%
amssymb,newlfont]{revtex}
\begin{document}
\draft
\preprint{
\begin{tabular}{r}
IASSNS-AST 97/54\\
hep-ph/9710251
\end{tabular}
}
\title{Neutrinos: past, present, future}
\author{S.M. Bilenky}
\address{Joint Institute for Nuclear Research, Dubna, Russia,\\
and\\
Institute for Advanced Study, Princeton, N.J. 08540}
\maketitle
\begin{abstract}
A general overview of neutrino physics is given.
The history
of the neutrino starting from Pauli and Fermi is briefly
presented.
Two-component neutrinos, the phenomenological
$V-A$ theory and the standard model are discussed.
The problem of neutrino 
masses and mixing is reviewed.
\end{abstract}

\pacs{Lecture presented at the
\emph{International School of Physics "Enrico 
Fermi"}, Varenna, Italy, 22 July -- 1 August 1997.}

\narrowtext

\section{Introduction}

Neutrinos  play
a special role in particle physics and astrophysics.
This is determined by the fact that
neutrinos have
only weak interactions.
Let me list some of the most important discoveries related to 
neutrinos.

\renewcommand{\labelenumi}{\arabic{enumi}.}
\begin{enumerate}

\item
In the 1950'
the electron 
neutrino was discovered
in the experiments of Reines and Cowan.

\item
In 1956 the parity non-conservation in $\beta$-decay 
was discovered (Wu \textit{et al.}).

\item
In 1957 it was proved that
the neutrino is a left-handed particle (Goldhaber \textit{et al.}).

\item
In the 1962 Brookhaven experiment of
Lederman, Steinberger, Schwarz \textit{et al.},
the muon neutrino was 
discovered.

\item
In 1973
in a neutrino experiment by the Gargamelle collaboration  
at CERN a new class of weak interactions
(neutral currents)
was discovered.

\item
In the 1980's in experiments on neutrino beams
at CERN and at Fermilab, the
quark structure of nucleon was established and investigated.

\item
In 1983 at CERN, the intermediate $W$ and $Z$ bosons were discovered
(UA1 and UA2 collaborations).

\item
In the 70's  solar neutrinos were detected in pioneering  experiment
by R.Davis.

\item
In 1987 neutrinos from Supernova 1987A were detected 
(Kamiokande, IMB, Baksan).

\item
In the 1990's in LEP experiments it was found that only three types
of light flavour neutrinos 
exist in nature.

\end{enumerate}

This is an impressive list of
discoveries, the importance of which 
for elementary particle physics and astrophysics
is difficult to overestimate.

\section{The Pauli hypothesis}

I will start with the history of the neutrino.
The neutrino 
hypothesis   
was put forward by Pauli in 1930. At that time protons 
and electrons were considered as elementary  particles,
and nuclei were considered as bound states of protons and 
electrons.
Such picture confronted with two fundamental problems:

\renewcommand{\labelenumi}{\Roman{enumi}.}
\begin{enumerate}

\item
The problem of $\beta$-spectrum.

Continuous $\beta$-spectra cannot be explained if 
$\beta$-decay is a transition of
one nucleus into a two-body final state with another nucleus and
electron.

\item
The problem of the spin of some nuclei.

The classical example is the nucleus $^7 N_{14}$.
If nuclei are bound state 
of protons and electrons,
this nucleus must be a bound state
of 14 protons 
and
7 electrons and must have a half-integer spin.
From the experimental data it followed that $^7 N_{14}$ nuclei are
particles with integral spin.

\end{enumerate}

To solve these problems Pauli assumed that in addition to
$p$ and $e$ there exist
a new elementary particle (which he called the neutron)
with a spin 1/2, equal to zero electric charge, mass less than the
electron mass and an interaction much weaker than photon interaction.
He assumed that "neutrons" are constituents of nuclei
(thus solving the problem of the spin of $^7 N_{14}$ and other nuclei)
and that $\beta$-decay 
is a three-body decay with a nucleus, an electron and
a "neutron", (which is not detected in the experiment)
in final state.

In 1932 the neutron was discovered by Chadwick and all
nuclear data (including the spin of $^7 N_{14}$ and other nuclei)
can be naturally explained under the assumption that 
nuclei are bound states of protons and neutrons. The problem of
$\beta$-decay remained unsolved.

\section{Fermi theory of $\beta $-decay}

The first theory of $\beta$-decay was proposed by Fermi in 1934.
This theory was based on the assumption that nuclei are bound  
states of protons  and neutrons. Fermi
 assumed  that the light Pauli particle 
(which he named the neutrino) exists and is produced in
$\beta$-decay together with an electron in the process
\begin{eqnarray}
n \rightarrow p + e^{-} + \nu
\end{eqnarray}

The first Hamiltonian of $\beta$-decay 
was build by Fermi in analogy with the Hamiltonian
of electromagnetic interaction
\begin{eqnarray}
\mathcal{H}_{I}^{em}=e{\bar p}{\gamma}_\alpha p~A^{\alpha} 
\end{eqnarray}
that describes the transition
\begin{eqnarray}
p \rightarrow p~+~\gamma
\end{eqnarray}

Fermi assumed that the Hamiltonian of
the process (1) has a similar to (2) vector form:
\begin{eqnarray}
\mathcal{H}_I ^{\beta}= G_{F} \bar{p} \gamma_{\alpha} n~ \bar{e} 
\gamma^{\alpha} 
\nu~+~h.c.
\end{eqnarray}
where $G_F$ is the interaction constant.

Let us stress an important difference between the Hamiltonians
(2) and (4).
The interaction constant $ G_F$ has dimension $ M^{-2}$
( units $\hbar=c=1$, $M$ is a mass), while an electric charge $e$ is 
a dimensionless
quantity. This is connected to the fact that
the Fermi interaction (4) is a four-fermion interaction
and the electromagnetic
interaction (2) is the interaction of the pair of fermions
with a boson. 

The experiments on the investigation of $\beta$-decay, 
that were done after the Fermi theory appeared, showed
that the Fermi interaction is not enough to explain the data.
In 1936 Gamow and Teller 
proposed the following most general four-fermion
Hamiltonian of $\beta$-decay:
\begin{eqnarray}
\mathcal{H}_I ^{\beta}=\sum_i G_i \bar{p} O_i n ~\bar{e} O^i \nu~+~ 
h.c.
\end{eqnarray}
Here $ O_i\rightarrow 
1$(scalar),~$\gamma_{\alpha}$ (vector), $\sigma_{\alpha 
\beta}$ (tensor), $\gamma_\alpha 
\gamma_5$ (axial), $\gamma_5$ (pseudoscalar). 
In the Hamiltonian (5) five arbitrary constants enter and
there was general belief that the number of fundamental constants
in the $\beta$-decay Hamiltonian must be much less.
During many years 
the strategy of $\beta$-decay
experiments was to find dominant variants of
$\beta$-decay.
The situation remained uncertain up to 1956.

\section{Non-conservation of parity}

In 1956 
it was discovered that parity is not conserved 
in $\beta$-decay (Wu \textit{et al.}). This discovery completely 
changed 
our understanding of $\beta$-decay and neutrino.

In the experiment of Wu \textit{et al.} the dependence of
the probability of $\beta$-decay of polarized $^{60} Co$ on the angle
between the directions of the vector of polarization and
the electron momentum
was measured. In the case of non-conservation of
parity the
 probability of $\beta$-decay of 
polarized
nuclei is given by the following general expression:
\begin{eqnarray}
dW_{\vec{P}}(\vec{k} )=
dW_0 (1~+~\alpha \vec{P} \vec{k}),
\end{eqnarray}
where $\vec{P}$ is the polarization vector and 
$ \vec{k}=\frac{\vec{p}}{|\vec{p}|}$
($\vec{p}$ is electron momentum). If parity is 
conserved, in this case
\begin{eqnarray}
dW_{\vec{P}}(\vec{k})=
dW_{\vec{P}}(-\vec{k})
\end{eqnarray}
and the asymmetry parameter $\alpha$ is equal to zero
(the pseudoscalar $\vec{P} \vec{k}$ cannot enter 
in the expression
for the probability of the decay in the case of parity conservation).
In the experiment of Wu \textit{et al.} it was found that $\alpha 
\simeq-0.7$.

The idea of non-conservation of parity in
a weak interaction was put forward by Lee and Yang before
the Wu \textit{et al.} experiment. The
Hamiltonian of $\beta$-decay 
that they proposed
was a direct generalization of the four-fermion
Fermi-Gamow-Teller Hamiltonian:
\begin{eqnarray}
\mathcal{H}_I^{\beta}=\sum_i \bar{p} O_i n~ \bar{e} O^i
(G_i~+~G_i ^{'}\gamma _5)\nu~+~h.c.
\end{eqnarray}
This Hamiltonian is the most general four-fermion Hamiltonian
in the case of non-conservation of parity.
It
is characterized by 10 (!) arbitrary 
constants $G_i$ and $G_i^{'}~(i=S,~V,~T,~A,~P~)$.

\section{Two-component neutrino}

In 1957-58 two fundamental stages completely changed the field
and brought us to the correct effective theory of $\beta$-decay 
and
other weak processes.
The first step was done by
Landau, Lee and Yang, and Salam.
They proposed the two-component neutrino theory.

Any fermion field $\psi$ can be presented as the sum of
left-handed $\psi_L$ and right-handed $\psi_R $ components
\begin{eqnarray}
\psi = \psi_L~+~\psi_R,
\end{eqnarray}
where 
\begin{eqnarray}
\psi_{L,R}=\frac{1 \mp \gamma_5}{2} \psi
\end{eqnarray}
In the case of massless neutrinos, 
$\nu_L (\nu_R)$
is the field of neutrino with left (right) helicity and
antineutrino with right (left) helicity.

According to the two-component neutrino theory, the neutrino is a massless
left-handed (or right-handed) particle, \textit{i.e.} in the 
Hamiltonian of 
weak interaction only left-handed $\nu_L$
(or right-handed $\nu_R$) neutrino field enters.
Thus, in the case of a two-component neutrino
theory $G_{i}^{'}=-G_{i}$ $(G_{i}^{'} = G_i)$
and non-conservation of parity in $\beta$-decay
is maximal.

In 1957 the two-component theory was confirmed by the experiment
of Goldhaber \textit{et al.}. In this 
experiment
by
the measurement of the polarization of 
$\gamma$-quanta in the process
\begin{eqnarray}
\begin{array}{rcl} \displaystyle
e^- + Eu \to \nu +
\null & \null \displaystyle
Sm^*
\null & \null \displaystyle
\\
& \null \scriptstyle \Downarrow \null &
\\ \displaystyle
\null & \null \displaystyle
Sm
\null & \null \displaystyle
+ \gamma
\end{array}
\end{eqnarray}
the helicity of neutrinos was determined.
It was found that the neutrino
is a left-handed particle.

\section{$V-A$ current $\times$ current theory}

The next decisive stage in the 
 construction of an effective theory of weak interaction
was done by Feynman and Gell-Mann, Marshak and Soudarshan.
These authors assumed that in the Hamiltonian of the weak interaction
 only left-handed components of fields enter.  
Taking into account that
\begin{eqnarray}
{\bar e}_L~(~1~;~\sigma_{\alpha \beta}~;~\gamma_5~)~\nu_L=0
\end{eqnarray}
and that
\begin{eqnarray}
{\bar e}_L \gamma_{\alpha} \gamma_5 \nu_L=
-{\bar e}_L \gamma_{\alpha} \nu_L
\,,
\end{eqnarray}
for the Hamiltonian of $\beta$-decay from (8) they 
obtained
the following expression
\begin{eqnarray}
\mathcal{H}_I ^{\beta}=\frac {G_F}{\sqrt{2}} 4 {\bar p}_L \gamma_{\alpha}
n_L ~{\bar e}_L \gamma^{\alpha} \nu_L~+~h.c.
\end{eqnarray}
This interaction is characterized by  one
fundamental constant $ G_F$
and differs from the Fermi interaction (4) 
only in the change of all fields by left-handed fields.
From all available experimental data it follows that
the Hamiltonian (14) is the correct effective Hamiltonian
of $\beta$-decay.

The other process in which a neutrino is produced
is $\mu$-capture
\begin{eqnarray}
\mu ^-~+~p \rightarrow \nu~+~n
\end{eqnarray}
B. Pontecorvo was the first to notice
in the 1950's that
the constant that characterize this process is the Fermi constant.
He assumed that
the weak interaction is a universal interaction which includes
the pairs 
($e,~\nu$) and
($\mu,~\nu$). The theory of Feynman and Gell-Mann 
is a universal theory of the weak interaction. Their Hamiltonian
describes not only $\beta$-decay and $\mu$-capture, but also $\mu$-decay,
the process in which two neutrinos are produced
\begin{eqnarray}
\mu ^+ \rightarrow e^+~+~\nu~+~{\bar \nu}.
\end{eqnarray}

Feynman and Gell-Mann introduced a $V-A$ weak current
\begin{eqnarray}
j_{\alpha}=2[{\bar \nu}_{eL} \gamma_{\alpha}e_L~+~{\bar \nu}_{\mu L}
\gamma_{\alpha} \mu_L~+~{\bar p}_L \gamma_{\alpha}n_L]
\end{eqnarray}
and assumed that the Hamiltonian of the weak interaction
has the current $\times$ current form
\begin{eqnarray}
\mathcal{H}_I=\frac{G_F}{\sqrt 2} j_{\alpha}j^{\alpha \dagger}
\end{eqnarray}
The non-diagonal terms of (18) are Hamiltonians
of $\beta$-decay, $\mu$-capture, $\mu$-decay and other
connected processes (like $ \nu_e n \to e^- p $, $\ldots$).
There 
are also in (18)
diagonal terms as
\begin{eqnarray}
\mathcal{H}_I^d=
\frac{G_F}{\sqrt{2}} \,
4 \,
\bar\nu_{eL} \gamma_{\alpha} \nu_{eL} \,
\bar{e}_{L} \gamma^{\alpha} e_L
\,.
\end{eqnarray}
Thus, the current $\times$ current theory
predicted new weak processes such as
\begin{eqnarray}
{\bar \nu}_e~+~e \rightarrow {\bar \nu}_e~+~e
\end{eqnarray}
Let us notice that this process 
with reactor antineutrinos was 
observed many years later by Reines \textit{et al.}. 
The measured cross section  was in agreement
with the standard model, that includes the interaction (19)
as well as the additional (neutral current) interaction.

\section{$\nu_\mu$ and $\nu_e$ are different particles}

In (17) we denoted the field of the neutrino
that enter
 in the Hamiltonian together with the field of electron (muon)
as $\nu_e (\nu_\mu)$.
 Are electron and muon neutrinos different or are they the same 
particles?
In 1959 B. Pontecorvo proposed an experiment that could allow to answer this
question. The idea of the experiment was the following.
According to the $V-A$ theory, the decay of a charged pion 
into an electron and a neutrino is strongly suppressed.
This consequence of the theory
was beautifully confirmed by the CERN experiment of Fidecaro \textit{et 
al.}.
Thus,
charged pions decay mainly into a muons and a muon neutrinos.
If we produce the beam of
pions
and give pions the possibility to decay, a practically pure
 beam of muon neutrinos can be produced.
In a neutrino detector 
 only muons would be observed if $\nu_\mu$  and $\nu_e$ are different 
particles.
If $\nu_\mu$ and $\nu_e$ are identical particles
due to 
$\mu-e$ universality an
equal number of muons and electrons will be produced.

The experiment proposed by Pontecorvo was done at Brookhaven by
Lederman, Steinberger, Schwarz \textit{et al.} in 1962. 
The Brookhaven experiment was the first
experiment with accelerator neutrinos. 
It was proven that $\nu_\mu$ and $\nu_e$ are different particles.

What are the quantum numbers that distinguish muon and electron neutrinos?
Let us introduce the electron $L_e$ and muon $L_\mu$ lepton numbers 
in such a way that $ L_e=1$, $L_\mu=0$ for 
$\nu_e$, $e^-$ and 
$ L_e=0$, $L_\mu=1$ for $\nu_\mu$, $\mu^-$.
The data of the Brookhaven experiment was in 
agreement
with the assumption that the total electron lepton number
and the total muon lepton number are conserved
\begin{eqnarray}
\sum L_e = \mbox{const}
\,,
\quad
\sum L_\mu = \mbox{const}
\,.
\end{eqnarray}

\section{Cabibbo and GIM currents}

Up to now we have not discussed decays of strange particles.
Strange particles  were included in the current $\times$ current scheme
by N. Cabibbo in 1962.

After many years of investigation 
of semileptonic decays of strange particles 
\begin{eqnarray}
K^+ \rightarrow {\pi^0 \mu^+ \nu_\mu},~
\Lambda \rightarrow {p e^- {\bar\nu}_e},~
\Sigma^- \rightarrow{n e^- {\bar\nu}_e}~,
\ldots
\end{eqnarray}
three phenomenological rules 
were formulated:

\renewcommand{\labelenumi}{\Roman{enumi}.}
\begin{enumerate}

\item
In semileptonic decays the strangeness is changed
by one: $|\Delta S|=1$.

\item
The rule $\Delta Q= \Delta S$ is satisfied, where
$\Delta Q= Q_f-Q_i$, $\Delta S =S_f-S_i 
$ and $Q_f$ $(Q_i)$ and $S_f$ $(S_i )$ are the
initial (final)
electric charge and strangeness of hadrons.

\item
The decays of strange particles are suppressed with respect 
to
decays of non-strange
particles.

\end{enumerate}

In order to include strange particles in the $V-A$ scheme,
Cabibbo assumed that the weak current is the combination of the
components of an SU(3) current.
We can easily construct the Cabibbo current if we assume that  
fields 
of $ u(Q=2/3,S=0)$, $d(Q=-1/3,S=0)$ and $s(Q=-1/3,S=-1)$ quarks 
enter in the weak current.
Let us stress that this is a new point of view:
this assumption means that the weak interaction is the interaction
of leptons and quarks. Let us accept the Feynman-Gell-Mann
conjecture and assume that only the left-handed components of fields
enter in the current. There are only the following two 
quark terms that, 
like the lepton terms in (17),
change the electric charge by one:
\begin{eqnarray}
{\bar u}_L \gamma _{\alpha} d_L~~ \mbox{and}~~ {\bar u}_L \gamma_{\alpha} s_L
\end{eqnarray}
The first term does not change the strangeness.
The second term changes the strangeness by one.
It is obvious that in the framework of the current $\times$  current scheme 
this term 
provides the fulfillment of the $ |\Delta S|=1$ 
and $\Delta Q= \Delta S$ rules
for semileptonic decays of strange particles.
In order that rule III be satisfied, Cabibbo introduced an 
angle 
$\theta_C $
and assumed that the quark current has the form
\begin{eqnarray}
j^C_{\alpha}=2[\cos 
{\theta_C} {\bar u}_L \gamma_{\alpha} d_L +
\sin{\theta _C} {\bar u}_L \gamma_{\alpha} s_L]
\,.
\end{eqnarray}
He showed that with the help of (24) it is possible to describe the
data.
For the parameter $ \sin{\theta_C}$ he found the value 
$ \sin{\theta_C} \simeq 0.2$.

Now the weak current can be written in the form
\begin{eqnarray}
j_\alpha=2 [{\bar\nu}_{eL} \gamma_{\alpha} e_L +{\bar\nu}_{\mu L}
\gamma_{\alpha} \mu_L + {\bar u}_L \gamma_{\alpha} d_L^{c} ],
\end{eqnarray}
where
\begin{eqnarray}
d_L^{c}=\cos{\theta_C} d_L + \sin{\theta _C} s_L
\;. 
\end{eqnarray}
As seen from (25) the lepton and quark terms have 
similar structure. However, 
there is an asymmetry between the lepton and quark parts of the current:
 there
are two lepton terms and one quark term. 

In 1970 strong arguments appeared
 in favour of an additional quark term in the weak
current (Glashow, Iliopoulos, Maiani).
It was necessary to
introduce an additional quark term in order
to suppress decays like 
 $ K^+ \rightarrow \pi ^+ \nu {\bar \nu}$, in which 
the strangeness of hadrons is changed and the charge is not changed
(such decays were not observed in the
experiments).

In order to introduce an additional term in the current 
it was necessary to assume that
a new quark with the charge $2/3$  exist.
This quark was called charmed (c). Glashow, Iliopoulos and Maiani 
assumed 
that
the additional term in the current has the form
\begin{eqnarray}
j_{\alpha}^{\mathrm{GIM}} =2{\bar c}_L \gamma_{\alpha}s_L^{c}
\,,
\end{eqnarray}
where
\begin{eqnarray}
s^{c}_L=-\sin{\theta_C} d_L+ \cos{\theta_C} s_L
\,.
\end{eqnarray}
is the combination of the fields of $s$ and $d$ quarks
orthogonal to the Cabibbo combination (26).
The weak current that includes the Cabibbo and GIM currents
has the form
\begin{eqnarray}
j_{\alpha}=2\left[\sum_{l=e,\mu} {\bar\nu}_{lL} \gamma_{\alpha} l_L + 
{\bar u}_L \gamma_{\alpha} d^{c}_L + {\bar c}_L 
\gamma_{\alpha}s^{c}_{L}\right]
\end{eqnarray} 
The charmed particles were discovered in 1975. The
investigation of the decays of charmed particles and neutrino
processes fully confirmed the GIM hypothesis.

\section{The standard theory of electroweak interactions}

Up to now we have discussed only a four-fermion weak
interaction.
 In 1938 O. Klein assumed
that there exist a charged heavy intermediate 
vector boson $W$ and that fundamental weak interaction
 is the interaction of
two fermions and a $W$-boson (like the electromagnetic 
interaction).
From this point of view, the weak processes at small $Q^2 $
(momentum transfer 
squared) such as 
$\beta$-decay,
are second order processes
with a virtual $W$-boson.

In the framework of the $V-A$ theory two alternative theories
were considered:

\renewcommand{\labelenumi}{\Roman{enumi}.}
\begin{enumerate}

\item
The theory of the four-fermion weak interaction.

\item
The theory with the intermediate
vector boson.

\end{enumerate}

The Hamiltonian of the four-fermion interaction is given
by (18). The Lagrangian of the interaction of fermions with
vector boson is given by
\begin{eqnarray}
\mathcal{L}_I=-\frac{g}{2 \sqrt 2} j_{\alpha} W^{\alpha} + h.c.
\end{eqnarray}
where $g$ is a dimensionless coupling constant.

From the point of the view of 
the intermediate vector theory the current-current
Hamiltonian (18)
is an effective Hamiltonian that describes second order weak processes
at $Q^2 << M_W ^2$ ($M_W$ is the mass of the $W$-boson).
The Fermi constant $G_F$ is connected with the constant $g$
and the mass $M_W $ by the relation
\begin{eqnarray}
\frac{G_F}{\sqrt 2}=\frac{g^2}{8 M_W^2}
\end{eqnarray}
and naturally has dimension $M^{-2}$.

The four-fermion theory,
as well as the theory with the intermediate vector boson
in the lowest order of the perturbation theory describe 
numerous
experimental data. However both theories 
were unrenormalizable theories. 

Over the
years
there were many attempts to build a renormalizable theory of weak 
interactions.
Success was achieved by way of the unification
of weak and electromagnetic interactions in a unified electroweak
interaction (Glashow 1961, Weinberg 1968, Salam 1968).

In order to unify weak and electromagnetic interactions
it is necessary to assume that a vector intermediate $W$-boson
exists. The unification is based upon local Yang-Mills gauge 
invariance.
In the case of the simplest SU(2) local gauge invariance with the doublets
of the left-handed fermion fields $\psi_{aL}$,
vector gauge fields $A_{\alpha}^i (i=1,2,3) $
are fields
of charged and neutral particles.
The interaction Lagrangian is given by
\begin{eqnarray}
\mathcal{L}_I= -g \sum_{i} j_{\alpha}^{i} A^{i\alpha}
\end{eqnarray}
where $j_{\alpha}^{i} =\sum_{a} {\bar \psi}_{aL}\gamma_{\alpha} 
\frac{1}{2} \tau_i \psi_{aL} $ 
is an isovector current. The Lagrangian (32)
can be written in the form
\begin{eqnarray}
\mathcal{L}_I =-\frac{g}{2\sqrt 2} j_{\alpha} W^{\alpha} + h.c. - 
g j_{\alpha}^3 A^{3\alpha}
\end{eqnarray}
where  $j_{\alpha}=2(j_{\alpha}^{1} + ij^2_{\alpha})$ is the charged 
current,
$ W_{\alpha} =\frac{1}{\sqrt 2}(A^{1}_{\alpha} -
i A^2_{\alpha})$ is the field of
charged vector particles and $ A^3_{\alpha}$ is the field 
of
neutral vector particles. If the fermion doublets are
chosen as
\begin{eqnarray}
\psi _{lL}=\pmatrix{\nu_{lL}^{'}\cr l_L^{'}\cr},~l=e,\mu,\tau,~
\psi _{1L}=\pmatrix{u_L^{'}\cr d_L^{'}\cr},~ 
\psi _{2L}=\pmatrix{c_L^{'}\cr s_L^{'}\cr},~
\psi _{3L}=\pmatrix{t_L^{'}\cr b_L^{'}\cr},
\end{eqnarray}
for the charged current we have the expression
\begin{eqnarray}
j_{\alpha}=2\left[\sum_l{\bar \nu}^{'}_{lL} \gamma_{\alpha}l^{'}_L+
{\bar u}^{'}_L \gamma_{\alpha}d^{'}_L+
{\bar c}^{'}_L \gamma_{\alpha}s^{'}_L+
{\bar t}^{'}_L \gamma_{\alpha}b^{'}_L \right]
\,,
\end{eqnarray}
that is similar to the expression (29) for the
phenomenological
charged current 
(in accordance with the existing data, additional lepton and quark 
terms 
are taken into account here).

The last term of the Lagrangian (33) describes the interaction of 
fermions with neutral vector bosons. This term includes the neutrino 
fields 
and
does not conserve parity; it is not the Lagrangian of
the electromagnetic
interaction.

In order to unify weak and electromagnetic interactions
we must enlarge the symmetry group and introduce an additional gauge
interaction with a neutral vector field. The interaction with
the charged vector field must be retained.
The minimal enlargement is a local gauge SU(2)$\times$U(1) group.
 In the case
of this group the gauge interaction has the form
\begin{eqnarray}
\mathcal{L}_I= -g{\vec{j}_{\alpha}} {\vec{A}}^{\alpha}-
g^{'}\frac{1}{2} j_{\alpha}^y B^{\alpha},
\end{eqnarray}
where $B^{\alpha}$ is a U(1) gauge field and $g^{'}$ is the
dimensionless
coupling constant.

 If the arbitrary U(1) constants
 are chosen in such a way that the Gell-Mann--Nishijima
rule $Q= I_3 + \frac{1}{ 2} y$ is satisfied, in this case
current $\frac{1}{2} j_{\alpha}^y$ is given by
\begin{eqnarray}
{1\over 2} j_{\alpha}^y = j_{\alpha} ^{em} - j_{\alpha}^3
\end{eqnarray}
where $j^{em}_{\alpha} $ is electromagnetic current.

The Lagrangian of the interaction of quarks and leptons
with neutral vector fields is equal to 
\begin{eqnarray}
\mathcal{L}_I^0=-gj^3_{\alpha}
 A^{3\alpha} - g^{'}(j_{\alpha}^{em}-j_{\alpha}^3)B^{\alpha}=
-ej_{\alpha}^{em} A^{\alpha}
-\frac{g}{2\cos{\theta_W}}j_{\alpha}^0 Z^{\alpha}
\end{eqnarray}
Here $g^{'}=g\tan{\theta_W}$, $g\sin{\theta_W}=e$,
the fields
$Z^{\alpha}$ and $ A^{\alpha}$ are connected with 
$ A^{3\alpha}$ and $B^{\alpha}$
by the relations
\begin{eqnarray}
Z^{\alpha}_{\alpha}=\cos{\theta_W} A^{3\alpha}-
\sin{\theta_W} B^{\alpha}_{\alpha}
\nonumber\\
 A^{\alpha}=\sin{\theta_W} A^{3\alpha}+\cos{\theta_W} 
B^{\alpha}
\end{eqnarray}
and
\begin{eqnarray}
j_{\alpha}^0=2j^3_{\alpha}-2\sin^2{\theta_W} j^{em}_{\alpha}
\end{eqnarray}
The first term of the expression (38)
is the Lagrangian of the electromagnetic interaction, the second term 
is
the new neutral current interaction of fermions and vector bosons.

Thus, if weak and electromagnetic interactions are
unified on the basis of local gauge symmetry,
in this case:
\renewcommand{\labelenumi}{\arabic{enumi}.}
\begin{enumerate}

\item
Not only the charged vector $W^{\pm}$-bosons but also the neutral 
vector 
$Z$-boson
must exist.

\item
The Lagrangian of the interaction of $Z$-bosons and fermions
has the form of the product of the $Z$-field and the neutral current 
$j^0_{\alpha}$,
which is a combination of the third component of isovector current
$j_{\alpha}^3$
and electromagnetic current $j_{\alpha}^{em}$; the only parameter 
which enters in the
neutral current is $\sin^2{\theta_W}$.

\item
The parameters of the theory are connected by the relation
$g\sin{\theta_W}=e$.

\end{enumerate} 

A local gauge invariance is an exact symmetry only if the masses of all 
particles
are equal to zero. Thus, in the real world, this symmetry must be 
broken.

The standard electroweak theory (standard model) is based on the 
Higgs 
mechanism
of spontaneous violation of the symmetry
which requires the 
existence of
a neutral scalar Higgs boson. 
As a result of the violation of the symmetry:

\renewcommand{\labelenumi}{\roman{enumi}.}
\begin{enumerate}

\item
All particles (perhaps with the exception of neutrinos) acquire 
masses.

\item
The primed fermion fields that enter in the doublets (34)
are connected with fermion fields with definite masses by
unitary transformation. For the quark fields we have
\begin{eqnarray}
\pmatrix{d_L^{'}\cr s_L^{'}\cr b_L^{'}\cr }=
V_L \pmatrix{d_L\cr s_L\cr b_L\cr},~~
\pmatrix{u_L^{'}\cr c_L^{'}\cr t_L^{'}\cr }=
U_L \pmatrix{u_L\cr c_L\cr t_L\cr }
\end{eqnarray}
where $u,~d,~...$ are fields of physical quarks.

As a result, we come to the following expression for the charged 
quark current
\begin{eqnarray}
j_{\alpha}=2\left[{\bar u}_L \gamma_{\alpha} d^c_L+
{\bar c}_L \gamma_{\alpha} s^c_L+
{\bar t}_L \gamma_{\alpha} b^c_L\right]
\end{eqnarray}
which, for the case of two generations, coincides with the
phenomenological current (29).
Here
\begin{eqnarray}
d^c_L=\sum_{q=d,s,b} V_{uq}q_L,~~~s^c_L=\sum_{q=d,s,b}V_{cq}q_L,~~~
b^c_L=\sum_{q=d,s,b}V_{tq}q_L,
\end{eqnarray}
where $ V=U^{\dagger}_L V_L$ is the unitary 
Cabibbo-Kobayashi-Maskawa mixing matrix.

\end{enumerate}

Neutral currents were discovered in the neutrino experiments at 
CERN 
in 1973. Their detailed investigation showed impressive agreement
of the standard model with experiment.

The relations (43) mean that the fields of quarks enter in the
charged current in mixed form. 
From numerous experimental data that are available today, 
we have rather detailed information
about the elements of the CKM mixing matrix $V$. What about neutrinos?
Are neutrino masses different from zero and
the fields of massive neutrinos
enter into charged current also in the mixed form?  The major aim
of present (and future) neutrino experiments is to answer
these fundamental questions.

\section{Massive and mixed neutrinos} 

We will finish this lecture with a short discussion of
the problem of
 neutrino mixing.
Let us stress first of all that there 
are more possibilities for the mixing of neutrinos 
than for the mixing of 
quarks.
This is connected with the fact that quarks are charged Dirac particles,
whereas for massive neutrinos 
there are two possibilities:

\renewcommand{\labelenumi}{\Roman{enumi}.}
\begin{enumerate}

\item
Neutrinos can be Dirac particles. In this case the total lepton number
$ L=L_e+L_{\mu}+L_{\tau}$ is conserved and neutrinos and 
antineutrinos
have opposite lepton numbers.
For the neutrino mixing in this case
we have
\begin{eqnarray}
\nu_{lL}=\sum_{k=1}^3 U_{lk}\nu_{kL} ~~~~l=e,\mu,\tau
\end{eqnarray}
where $U^{\dagger}U=1$ and $\nu_k$ is the field of the neutrino with 
mass $ m_k$.

\item
Neutrinos can be truly neutral Majorana particles.
In this case there are no conserved lepton numbers, 
and for neutrino mixing we have
\begin{eqnarray}
\nu_{lL}=\sum_{k=1}^n U_{lk}\chi_{kL}
\end{eqnarray}
where $\chi_k=\chi_k^c=C{\bar \chi}^{T}_k$ 
is the field of the Majorana neutrino with mass $ m_k$
($C$ is the charge conjugation matrix).

\end{enumerate} 

The number of massive neutrinos in (45) depends
upon the model. If only left-handed components
$ \nu_{lL}$ enter in the neutrino mass term, in this case
$n=3$. If left-handed
$ \nu_{lL}$ and right-handed $\nu_{lR}$ fields
enter in the neutrino mass term, in this case $n=6$.

Note that Dirac neutrino masses can be generated
by the standard Higgs mechanism. The Majorana neutrino masses
can be generated only in the framework of models beyond the standard
model.

The existing models cannot allow to predict the values of the neutrino 
masses.
There exist, however, a rather general mechanism of neutrino mass 
generation
that can explain the smallness of neutrino masses with respect to  
the 
masses of all the other fundamental fermions.
This is the so-called see-saw 
mechanism.
If we assume that ,
 due to the presence of a right-handed 
Majorana mass term,
lepton numbers are violated at a scale $M $ that is much larger then the
fermion masses, 
in this case
for the mass of the light neutrino in each generation
we have the see-saw formula
\begin{eqnarray}
m_k\simeq \frac{({m^k_F})^2}{M_k},~~~k=1,2,3.
\end{eqnarray}
Here $m^k_F~$ is the mass of the up-quark or charged lepton and
$M_k \gg m^k_F$. Let us stress that if the neutrino masses
are generated by see-saw mechanism,
in this case:

\renewcommand{\labelenumi}{\arabic{enumi}.}
\begin{enumerate}

\item
Massive neutrinos are Majorana particles.

\item
There is a hierarchy of neutrino masses:
$m_1\ll m_2\ll m_3$.

\end{enumerate}

At the moment  the problem of neutrino masses and mixing
is investigated in different experiments.
There are three experimental methods that allow to reveal
the effects of neutrino masses and mixing.

\subsection{Precise measurement of high energy part of beta-spectrum}

The classical decay in which  neutrino mass is measured is the $\beta$-decay 
of $^3H$
\begin{eqnarray}
^3H \rightarrow{^3He+e^-+{\bar\nu}_e}
\end{eqnarray}
This is the superallowed transition and
the electron spectrum is determined by the phase factor
\begin{eqnarray}
{dN\over{dT}}=CpE(Q-T)\sqrt{{(Q-T)}^2-m^2_{\nu}}\,F(E),
\end{eqnarray}
where $E$ and $p$ are the electron energy 
and momentum, $ T=E -m_e$, $Q\simeq 18.6 $ keV is the
energy release, $F(E)$ is the Fermi function that describes 
the Coulomb 
interaction 
of the final particles, and $m_\nu$ is the neutrino mass.
 In the real spectrum it is necessary to take into 
account molecular effects, spectrometer resolution, 
background, and so on.

From the measured spectrum the Kurie function 
\begin{eqnarray} 
K(T) = \sqrt{ (Q-T) \sqrt{ (Q-T)^2 - m^2_\nu } }
\end{eqnarray}
can be obtained. If $m_\nu =0$,
in this case
$T_{max}=Q$ and $K(T)=Q-T$. If $m_\nu \not=0$, in this case
$T_{max}=Q-m_\nu$
and a deviation of $K(T)$ from a straight line will be 
observed
near the end point of the spectrum.

No indications in favour of non-zero neutrino masses were found in 
tritium 
experiments.
From the data of recent experiments the following 
upper 
bounds
were found:
\begin{eqnarray}
m_{\nu}<3.9~eV\quad\hbox{(Troizk)},\quad 
m_{\nu}<5.6~ eV\quad\hbox{(Mainz)}.
\end{eqnarray}

\subsection{Search for neutrinoless double-$\beta$ decay}

There are many experiments in which neutrinoless double-beta decay
\begin{eqnarray}
(A,Z)\rightarrow {(A,Z+2)+e^-+e^-}
\end{eqnarray}
of different even-even nuclei is 
searched for. 
In  process (51) the total lepton number is not conserved.
Thus neutrinoless double-beta decay
is possible only if neutrinos are massive Majorana particles.

Process
(51) is of the second order 
in  weak interaction
with virtual neutrinos.
The matrix element of the process is proportional to
\begin{eqnarray}
<m>=\sum{U^2_{ek}m_k}
\end{eqnarray}
($U^2_{ek}$ is due to two vertices and $m_k$ is due to 
the the propagator of left-handed neutrino fields).
 No indications in favour of neutrinoless double-beta decay
were found in experiments up to present time. In the $^{76}Ge$ 
experiment of
the Heidelberg-Moscow collaboration it was found that
\begin{eqnarray}
T_{\frac{1}{2}}\geq 1.1\cdot{10^{25}}y
\end{eqnarray}
From this data it follows that $ |<m>|\leq{(0.5-1.1)eV}$.
Let us notice that in the nearest years
sensitivity of $|<m>| \simeq{10^{-1}} eV$ will be reached
(NEMO, Heidelberg-Moscow and other experiments).

\subsection{Neutrino oscillations}

Neutrino oscillations 
were first considered by B. Pontecorvo in 1958. From
the point of view of quantum mechanics, neutrino oscillations are
similar to the famous 
$ K^0 \leftrightarrows \bar{K}^0 $ 
oscillations.
If there is neutrino mixing, the state vector 
$|\nu_l\rangle$ of the flavour neutrino $\nu_l$ with momentum $p$ 
is the coherent superposition 
of 
the states $|k\rangle$ of massive neutrinos with momentum $p$ and energy
$E_k=\sqrt{p^2+m^2_k} \simeq {p+\frac{m^2_k}{2p}}$ (for $ p \gg m_k $):
\begin{eqnarray}
| \nu_l\rangle=\sum_k{U^{\ast}_{lk}}|k\rangle
\,.
\end{eqnarray}
This basic relation is valid if the neutrino mass
 differences are so small
that, due to an uncertainty relation,
 it is not possible in weak interaction
to distinguish
one massive neutrino from the other one.  
If  at $t=0$ in weak decays neutrinos $\nu_l $
with momentum $p$ are produced,
at time $t$ the neutrino state vector is given by
\begin{eqnarray}
{|\nu_l\rangle}_t=\sum_k U^{\ast}_{lk}e^{-iE_kt}|k\rangle=
\sum_{l^{'}}|\nu_{l^{'}}\rangle\sum_k U_{l{'}k}e^{-iE_kt}U^{\ast}_{lk}
\end{eqnarray}
Thus ,if there is neutrino mixing, the beam of neutrinos
at some macroscopic distance from the source will be a superposition
of states of different flavour neutrinos.
For the probability of the transition $\nu_l\rightarrow \nu_{l{'}}$,
we have
\begin{eqnarray}
P(\nu_l\rightarrow \nu_{l{'}})={\left|\sum_k U_{l{'}k}
e^{-i\frac{\Delta m^2_{k1}L}{2p}}U^{\ast}_{lk}\right|}^2
\,.
\end{eqnarray}
Here $L$ is the source-detector distance 
and $\Delta{m^2_{k1}}=m_k^2-m_1^2$ (we made the usual 
assumption that $m_1<m_2<\ldots$).
Taking into account the unitarity of the mixing matrix, 
for the simplest case 
of oscillations
between two flavour neutrinos from (54) we have
\begin{eqnarray}
P(\nu_l\rightarrow \nu_{l{'}})
=\left|\delta_{l{'}l}+U_{l{'}2}U^{\ast}_{l2}
(e^{-i\frac{\Delta m^2L}{2p}}-1)\right|^2 ~, 
\end{eqnarray}
where $\Delta m^2=m^2_2-m^2_1$.
From this expression it is clear that for neutrino oscillations 
to be observed, it is necessary that 
\begin{eqnarray}
\Delta m^2 \gtrsim {\frac{E}{L}}
\,.
\end{eqnarray}
Here $L$ is the distance in meters, $E$ is neutrino energy in 
MeV and $\Delta m^2$ is the difference of neutrino masses squared 
in 
eV$^2$.
From (58) it follows that different neutrino facilities 
(accelerators, reactors, atmospheric neutrinos, sun)
allow us
to study neutrino oscillations in a wide range
of $\Delta m^2$, from $\Delta m^2 
\simeq 10 eV^2$ till $\Delta m^2 
\simeq 10^{-10} eV^2$.
For two oscillating neutrinos the mixing matrix is given by 
\begin{eqnarray}
\nonumber
U=\pmatrix{\cos\theta &\sin\theta\cr
-\sin\theta& -\cos\theta\cr}
\,,
\end{eqnarray}
where $\theta$ is the mixing angle.
From (57) for the transition probabilities we obtain the 
following standard expressions
\begin{eqnarray}
P(\nu_l\rightarrow \nu_{l{'}})=
\frac{1}{2}\sin^2{2\theta}(1-\cos{\frac{\Delta m^2L}{2p}})
\nonumber\\
P(\nu_l\rightarrow \nu_l)=1-P(\nu_l \rightarrow 
\nu_{l{'}})
\,,
\qquad
l^{'}\not=l
\end{eqnarray}

In many experiments with neutrinos from
reactors and accelerators
no indications in favour of neutrino oscillations 
were found.
There are, however,
 three experimental indications
that neutrinos are massive and mixed.
They were found in solar neutrino experiments,
in atmospheric neutrino experiments, 
and in the Los Alamos neutrino experiment.

Let us first discuss solar neutrinos. The energy of the sun 
is 
generated in 
the reactions of the thermonuclear $pp$ and CNO cycles.
From the thermodynamical point of view,
energy is generated in the
transition 
of four protons into $ ^4He$:
\begin{eqnarray}
4p \rightarrow{^4He+2e^++2\nu_e}
\,.
\end{eqnarray}
Thus, the generation of the energy of the sun is accompanied
by the emission of electron neutrinos.
The total flux of
neutrinos is connected to the Luminosity of the sun 
$L_\odot$ by the relation
\begin{eqnarray}
Q \sum_{i} \left(1-2 \frac{{\bar E}_i}{Q}\right) {\Phi_i}=
\frac{L_\odot}{2\pi R^2}
\,,
\end{eqnarray}
where $Q=26.7$ MeV is the energy release in the 
transition (60), $R$ is the sun-earth 
distance, $ \Phi_i$
is the total flux of neutrinos from the source $i$, and ${\bar 
E}_i$ is the average 
energy.
The most important sources of solar neutrinos are 
the following reactions:
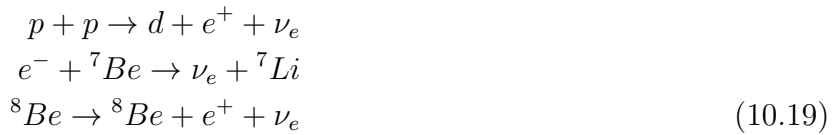
\begin{eqnarray}
p+p\rightarrow d+e^++\nu_e
\nonumber\\
e^-+ {^7Be}\rightarrow \nu_e+ {^7Li}
\nonumber\\
^8Be\rightarrow{^8Be+e^++\nu_e}
\end{eqnarray}
The first reaction 
is the main source of solar neutrinos. In this reaction
 neutrinos
with energy less than 0.42 MeV are produced. The
second reaction is a source of monochromatic neutrinos
with energy 0.86 MeV.
This reaction contributes about $10\%$ to the total flux of solar 
neutrinos. The third reaction contributes only about $ 10^{-4}$ to
the total flux. However, this reaction is the main source
of high energy solar neutrinos (up to 15 MeV).

At present the data of the
Homestake, GALLEX, SAGE, Kamiokande and Super-Kamiokande experiments
are available. 
These experiments, due to different
detection thresholds, allow us
to detect neutrinos from different sources:
the GALLEX and SAGE experiments allow us to detect neutrinos from 
all sources,
the Homestake experiment allows us to detect mainly $^8B$ 
neutrinos 
(about 
$90\%$) and 
$^7Be $ neutrinos; and in the experiment Kamiokande and 
Super-Kamiokande experiments
only $^8B$ neutrinos are detected.

In all experiments the observed event rate is significantly
smaller than the rate predicted by the standard solar model.
For example, in 
the Super-Kamiokande experiment
the detected  flux  is 
equal to
\begin{eqnarray}
\Phi_{det}=2.44\pm 0.06^{+0.25}_{-0.09}~ 10^6~cm^{-2}s^{-1}\,.
\end{eqnarray}
The expected flux of $^8B$ neutrinos is equal to
\begin{eqnarray}
\Phi_{exp}=6.62\pm 1~ 10^6~cm^{-2}s^{-1}\,.
\end{eqnarray}

All existing data can be explained
if, due to an enhancement of the neutrino mixing in matter
(MSW effect), 
there are transitions of
initial solar $\nu_e$  into neutrinos of other types
that are not detected by existing experiments.
For the oscillation parameters $\Delta m^2$ and $\sin^2{\theta}$
two possible  values were obtained
\begin{eqnarray}
\Delta m^2 \simeq 5\cdot 10^{-6} eV^2 & &\sin^2{\theta} 
\simeq 7\cdot 10^{-3}
\nonumber\\
& \hbox{ or}&
\nonumber\\
\Delta m^2 \simeq 2\cdot 10^{-5} eV^2 & &\sin^2{2\theta} \simeq 
0.8
\end{eqnarray}
Let us notice that existing data can be also explained by the
vacuum oscillations.

The present analysis of solar neutrino data
is based on the standard solar model. Let
us stress that the Super-Kamiokande experiment and
the future SNO experiment(in which solar neutrinos will be 
detected
by the observation of CC and NC reactions) will allow us to 
obtain model-
independent
information about neutrino masses and mixing.

The second indication in favour of neutrino mixing was obtained
in atmospheric neutrino experiments. Atmospheric neutrinos are
produced in decays of pions and kaons ($\pi(K) 
\rightarrow \mu\nu_\mu$, $\mu \rightarrow e\nu_e \nu_{\mu}$)
that are produced in the interaction of cosmic rays 
with the
atmosphere. The ratio of the muon and electron events can be predicted
with accuracy about $5\%$ (at relatively small energies
this ration is close to 2). In the Kamiokande, IMB and Soudan 
experiments
it was found that the ratio $R$ of the ratio of observed muon and electron 
events
to the predicted ratio is significantly less than one. This atmospheric 
neutrino anomaly was confirmed recently by the
Super-Kamiokande experiment. 
For the ratio of ratios $R$ in this experiment it was found
\begin{eqnarray}
R=0.635\pm 0.033\pm 0.053
\,.
\end{eqnarray}
The result obtained can be explained by 
$ \nu_{\mu} \to \nu_{\tau} $
oscillations. For the oscillation 
parameters
it was found
\begin{eqnarray}
\Delta m^2 \simeq 5\cdot 10^{-3} eV^2,
\qquad
\sin^2{2\theta} \simeq 1.
\end{eqnarray}

Several long-baseline
neutrino oscillation experiments
(KEK--Super-Kamiokande, Fermilab--Soudan, 
CERN--Gran-Sasso, CHOOZ and Palo Verde)
 that will allow us 
to investigate the "atmospheric neutrino range" of $\Delta m^2$
in different oscillation channels
are now under preparation.
 The first reactor long-baseline experiment CHOOZ has 
started recently. 

The third indication in favour of neutrino mixing
was found in the Los Alamos accelerator experiment.
Neutrinos in this experiment 
are produced in
decays
at rest of $\pi^+$ and $\mu^+$:
\begin{eqnarray}
\pi^+ \rightarrow \mu^+\nu_\mu,
\qquad
\mu^+ \rightarrow e^+\nu_e{\bar\nu}_\mu.
\end{eqnarray}
The LSND detector at the distance about 30 m from beam stop
is searching for electron antineutrinos
(by the observation of 
the process
${\bar\nu}_e p\rightarrow e^+n$).
It was found 22 such events.
The expected background is $4.6\pm0.6$ events.
If negative results of other experiments are taken into account
from the results of the LSND experiment
the following allowed range of oscillation 
parameters was obtained
\begin{eqnarray}
0.3<\Delta m^2\leq2 ~eV^2,~~10^{-3} \leq\sin^2{2\theta} \leq 4\times
10^{-2}
\end{eqnarray}
Indications in favour of relatively large values
of $\Delta m^2$ in $\nu_\mu \rightarrow \nu_e $ channel
 were obtained only in LSND experiment.
These data need confirmation from other experiments.
It is planned that another experiment KARMEN will reach 
sensitivity 
of LSND 
experiment
in about 2 years.

\section{Conclusion}

After  Pauli and Fermi 
neutrino
physics have done tremendous progress
(see, for example, the books
\cite{Bahcall,Boehm-Vogel,Bilenky,CWKim,Mohapatra-Pal,Winter}).
We know that three types of flavour left-handed neutrinos
exist in nature and we know interaction of neutrinos
with other particles. 
However, the problem of neutrino properties remains unsolved.
The key problem is the problem of the neutrino mass
(see, for example, the reviews \cite{BP78,BP87}).
Today we have different indications that neutrinos are massive
(see, for example, the Proceedings \cite{Neutrino96,TAUP97}).
All of them require further checks and confirmation.
We do not know what is the nature of massive neutrinos
(Dirac or Majorana) and how many massive neutrinos exist in nature.
We need to know the neutrino mixing matrix. The solution
of these problems could bring us to new physics
beyond the standard model. The solution of the neutrino mass problem
will be very important for astrophysics and in particular
 for the understanding of the dark matter 
problem.

The investigation of neutrino properties is the present and 
future of neutrino physics.

\acknowledgments
I would like to acknowledge support from Dyson Visiting
Professor Funds at the Institute for Advanced Study.


\begin{references}

\bibitem{Bahcall}
J.N. Bahcall,
\emph{Neutrino Physics and Astrophysics},
Cambridge University Press, 1989.

\bibitem{Boehm-Vogel}
F. Boehm and P. Vogel,
\emph{Physics of Massive Neutrinos},
Cambridge University Press, 1987.

\bibitem{Bilenky}
S.M. Bilenky,
\emph{Introduction to Feynman Diagrams
and Electroweak Interaction Physics},
Editions Frontiers, 1994.

\bibitem{CWKim}
C.W. Kim and A. Pevsner,
\emph{ Neutrinos in Physics and Astrophysics},
Contemporary Concepts in Physics, Vol.8,
Harwood Academic Press, Chur, Switzerland, 1993.

\bibitem{Mohapatra-Pal}
R.N. Mohapatra and P.B. Pal,
\emph{Massive Neutrinos in Physics and
Astrophysics},
World Scientific Lecture Notes in Physics, Vol.41,
World Scientific, Singapore, 1991.

\bibitem{Winter}
\emph{Neutrino Physics},
Edited by K. Winter,
Cambridge University Press, 1991.

\bibitem{BP78}
S.M. Bilenky and B. Pontecorvo,
Phys. Rep. \textbf{41}, 225 (1978).

\bibitem{BP87}
S.M. Bilenky and S.T. Petcov,
Rev. Mod. Phys. \textbf{59}, 671 (1987).

\bibitem{Neutrino96}
Proceedings of \emph{Neutrino 96},
Helsinki, June 1996.

\bibitem{TAUP97}
Proceedings of \emph{TAUP97},
Laboratori Nazionali del Gran Sasso, Assergi (Italy),
September 1997.

\end{references}
\end{document}